# Bivariate Causal Discovery and its Applications to Gene Expression and Imaging Data Analysis

Rong Jiao[1], Nan Lin[1], Zixin Hu[2], David A Bennett[3], Li Jin[2] and Momiao Xiong[2,1*]

[1]Department of Biostatistics and Data Science, The University of Texas School of Public Health, Houston, TX 77030, USA.

[2]Ministry of Education Key Laboratory of Contemporary Anthropology, School of Life Sciences, Fudan University, Shanghai, China.

**Running Title**: Bivariate Causal Discovery

*Address for correspondence and reprints: Dr. Momiao Xiong, Department of Biostatistics and Data Science, School of Public Health, The University of Texas Health Science Center at Houston, P.O. Box 20186, Houston, Texas 77225, (Phone): 713-500-9894, (Fax): 713-500-0900, E-mail: Momiao.Xiong@uth.tmc.edu.

**Abstract**

The mainstream of research in genetics, epigenetics and imaging data analysis focuses on statistical association or exploring statistical dependence between variables. Despite their significant progresses in genetic research, understanding the etiology and mechanism of complex phenotypes remains elusive. Using association analysis as a major analytical platform for the complex data analysis is a key issue that hampers the theoretic development of genomic science and its application in practice. Causal inference is an essential component for the discovery of mechanical relationships among complex phenotypes. Many researchers suggest making the transition from association to causation. Despite its fundamental role in science, engineering and biomedicine, the traditional methods for causal inference require at least three variables. However, quantitative genetic analysis such as QTL, eQTL, mQTL, and genomic-imaging data analysis requires exploring the causal relationships between two variables. This paper will focus on bivariate causal discovery. We will introduce independence of cause and mechanism (ICM) as a basic principle for causal inference, algorithmic information theory and additive noise model (ANM) as major tools for bivariate causal discovery. Large-scale simulations will be performed to evaluate the feasibility of the ANM for bivariate causal discovery. To further evaluate their performance for causal inference, the ANM will be applied to the construction of gene regulatory networks. Also, the ANM will be applied to trait-imaging data analysis to illustrate three scenarios: presence of both causation and association, presence of association while absence of causation, and presence of causation, while lack of association between two variables.

**Introduction**

Despite significant progress in dissecting the genetic architecture of complex diseases by association analysis, understanding the etiology and mechanism of complex diseases remains elusive. Using association analysis and machine learning systems that operate, almost exclusively, in a statistical, or model-free modes as a major analytic platform for genetic studies of complex diseases is a key issue that hampers the discovery of mechanisms underlying complex traits (Pearl 2018).

As an alternative to association analysis, causal inference may provide tools for unraveling principles underlying complex traits. Power of causal inference is its ability to predict effects of actions on the system (Mooij et al. 2016). Typical methods for unraveling cause-and-effect relationships are interventions and controlled experiments. Unfortunately, the experiments in human genetics are unethical and technically impossible. Next generation genomic, epigenomic, sensing and image technologies produce ever deeper multiple omic, physiological, imaging, environmental and phenotypic data with millions of features. These data are almost all "observational", which have not been randomized or otherwise experimentally controlled (Glymour 2015). In the past decades, a variety of statistical methods and computational algorithms for causal inference which attempt to abstract causal knowledge from purely observational data, referred to as causal discovery, have been developed (Zhang et al. 2018). Causal inference is one of the most useful tools developed in the past century. The classical causal inference theory explores conditional independence relationships in the data to discover causal structures. The PC algorithms and the fast causal inference (FCI) algorithms developed at Carnegie Mellon University by Peter Spirtes and Clark Glymour are often used for cause discovery (Le et al. 2016). Despite its fundamental role in science, engineering and biomedicine, the conditional independence-based classical causal inference methods can only identify the graph up to its Markov equivalence class, which consists of all DAGs satisfying the same conditional independence distributions via the causal Markov conditions (Nowzohour and Bühlmann 2016). For example, consider three simple DAGs: $x \rightarrow y \rightarrow z$, $x \leftarrow y \leftarrow z$ and $x \leftarrow y \rightarrow z$. Three variables $x, y$ and $z$ in all three DAGs satisfy the same causal Markov condition: $x$ and $z$ are independent, given $y$. This indicates that these three DAGs form a Markov equivalence class. However, these three DAGs represent three different causal relationships among variables $x, y$ and $z$, which prohibits unique causal identification. These non-unique causal solutions seriously limit their translational application.

In the past decade, causal inference theory is undergoing exciting and profound changes from discovering only up to the Markov equivalent class to identify unique causal structure (Peters et al. 2012; Peters and Bühlman, 2014). A class of powerful algorithms for finding a unique causal solution are based on properly defined functional causal models (FCMs). They include the linear, non-Gaussian, acyclic model (LiNGAM) (Zhang et al. 2018; Shimizu et al. 2006), the additive noise model (ANM) (Hoyer et al. 2009; Peters et al. 2014), and the post-nonlinear (PNL) causal model (Zhang and Hyvärinen 2009).

In genomic and epigenomic data analysis, we usually consider four types of associations: association of discrete variables (DNA variation) with continuous variables (phenotypes, gene expressions, methylations, imaging signals and physiological traits), association of continuous variables (expressions, methylations and imaging signals) with continuous variables (gene expressions, imaging signals, phenotypes and physiological traits), association of discrete variables (DNA variation) with binary trait (disease status) and association of continuous variables (gene expressions, methylations, phenotypes and imaging signals) with binary trait (disease status). All these four types of associations can be extended to four types of causations. This paper focuses on studying causal relationships between two continuous variables.

The many causal inference algorithms using observational data require that two variables being considered as cause-effect relationships are part of a larger set of observational variables (Mooij et al. 2016). Similar to genome-wide association studies where only two variables are considered, we mainly investigate bivariate causal discovery to infer cause-effect relationships between two observed variables. To simplify the cause discovery studies, we assume no selection bias, no feedback and no confounding. We first introduce the basic principle underlying the modern causal theory. It assumes that nature consists of autonomous and independent causal generating process modules and attempts to replace causal faithfulness by the assumption of Independence of Cause and Mechanism (ICM) (Peters et al. 2017; Besserve et al. 2017; Schölkopf et al. 2012; Janzing et al. 2010; Lemeire et al. 2012). Then, we will present ANM as a major tool for causal discovery between two continuous variables. We will investigate properties of ANM for causal discovery. Finally, the ANM will be applied to gene expression data to infer gene regulatory networks and longitudinal phenotype-imaging data to identify brain regions affected by intermediate phenotypes. A program for implementing the algorithm for bivariate causal discovery with two continuous variables can be downloaded from our website https://sph.uth.edu/research/centers/hgc/xiong/software.htm.

**The independence principle of cause and mechanism for causal inference**

The philosophical causal principle assumes that nature consists of independent, autonomous causal generating process modules (Peters et al. 2017; Shajarisales). In other words, causal generating processes of a system's variables are independent. If we consider two variables: cause $X$ and effect $Y$, then the mechanism that generates cause $X$ and the mechanism that generates effect $Y$ from the cause $X$ are independent. Or, the process that generates the effect $Y$ from the cause $X$ contains no information about the process that generates the cause $X$. In the probability setting, this indicates that the cause distribution $P(X)$ and the conditional distribution $P(Y|X)$ of $Y$ given $X$ are independent. Statistics provides definition of independence between two random variables, but provides no tools for defining independence between two distributions (Peters et al. 2017). Algorithmic information theory can offer notion and mathematical formulation of independence between two distributions or independence of mechanisms (Janzing et al. 2010; Parascandolo 2017).

Consider a universal Turing Machine $T$. For any binary string $s$, we define Kolmogorov complexity $K_T(s)$ as the length of the shortest program that generates $s$, denoted as $s^*$, using universal prefix Turing machine $T$ that outputs $s$ and then stops (Peters et al. 2017; Kolmogorov 1965). Therefore, we have $K_T(s) = |s^*|$, where $|.|$ denotes the number of bits of a binary string. Intuitively, the Kolmogorov complexity measures the minimal amount of information required to generate $s$ by any effective process. Similar to conditional probability, we can also define conditional Kolmogorov complexity. The conditional Kolmogorov complexity $K(t|s)$ of string $t$ given $s$, is defined as the length of the shortest program that can generate $t$ from $s$ and then stops. The Kolmogorov complexity $K(t, s)$ of the concatenation of two strings $t$ and $s$ is defined as the length of the shortest program that generate string $t's$ where $t'$ is the prefix code of $t$.

Now we introduce "additivity of complexity" property. It can be shown that (Grunwald and Vitanyi 2004):

$$K(t,s) = K(t) + K(s|t^*), \quad (1)$$

where $t^*$ denotes the first shortest prefix program that generates $t$ and then stops and is in general uncomputable.

Algorithmic mutual information is defined as

$$I(s:t) = K(s) - K(s|t^*). \quad (2)$$

Substituting $K(s|t^*)$ in equation (1) into equation (2), we obtain

$$I(s:t) \stackrel{\pm}{=} K(s) + K(t) - K(s,t), \quad (3)$$

where the symbol $\stackrel{\pm}{=}$ implies that the equation can hold for up to constants. Equation (3) states that that this information is symmetrical: $I(s:t) = I(t:s)$. Therefore, $I(s:t)$ is called algorithmic mutual information between $s$ and $t$. The algorithmic mutual information quantifies the amount of information two strings or objects have in common, or the amount of bits saved when compressing $s, t$ jointly rather than compressing $s, t$ independently.

Similar to mutual information $I(s;t) = 0$ between two random variables where mutual information of zero implies independence of two variables, the algorithmic mutual information of zero $I(s:t)$ indicates algorithmically independence of two distributions of random variables. We also can define algorithmic conditional mutual information as

$$I(s:t|z) \stackrel{\pm}{=} K(s|z) + K(t|z) - K(s,t|z). \quad (4)$$

In statistics, although dependence between two random variables can be measured, there are no measures to quantify dependence between two distributions. We use algorithmic mutual information to measure independence between two distributions which can be used to assess causal relationships between two variables. Consider two variables $X$ and $Y$ and assume $X$ causes $Y$ ($X \to Y$). Let the marginal distribution of cause $X$ and conditional distribution of effect $Y$ given $X$ be $P_X$ and $P_{Y|X}$, respectively. The independence of cause and mechanism (ICM) states that the distributions $P_X$ and $P_{Y|X}$ are independent and hence $P_X$ and $P_{Y|X}$ are algorithmically independent, which implies that their algorithmic mutual information should be equal to zero:

$$I(P_X : P_{Y|X}) \stackrel{+}{=} 0, \tag{5}$$

or, equivalently,

$$K(P_{X,Y}) \stackrel{+}{=} K(P_X) + K(P_{Y|X}). \tag{6}$$

In other words, distributions $P_X$ and $P_{Y|X}$ have no common information. If $X$ causes $Y$, then the conditional distribution $P_{Y|X}$ of the effect $Y$ given cause $X$ contains no information about cause $X$. Algorithmic mutual information is asymmetric. Thus, the algorithmic mutual information can be used to infer whether $X \to Y$ or $Y \to X$. If $I(P_X : P_{Y|X}) < I(P_Y : P_{X|Y})$ then $X \to Y$. Similarly, if $I(P_X : P_{Y|X}) > I(P_Y : P_{X|Y})$ then $Y \to X$. Cause and effect cannot be identified from their joint distribution. Cause and effect are asymmetric. The joint distribution is symmetric. It can be factorized to $P_{X,Y} = P_X P_{Y|X} = P_Y P_{X|Y}$.

This implies that the joint distribution $P_{X,Y}$ of two variables $X, Y$ is unable to infer whether $X \to Y$ or $Y \to X$. Peters et al. (2014) showed in Proposition 4.1 of their book that for every joint distribution $P_{X,Y}$ or $P_{Y,X}$ of real-valued variables $X$ and $Y$, there are nonlinear models:

$$Y = f_Y(X, N_Y), X \perp\!\!\!\perp N_Y$$

and

$$X = g_X(Y, N_X), Y \perp\!\!\!\perp N_X,$$

where $f_Y$ and $g_Y$ are functions and $N_Y$ and $N_X$ are real-valued noise variables. In supplementary note we provide the details that were omitted in the proof of Proposition 4.1 (Peters et al. 2014). This shows that to make a bivariate causal model identifiable, we must restrict the function class.

**Nonlinear additive noise models for bivariate causal discovery**

Assume no confounding, no selection bias and no feedback. Consider a bivariate additive noise model $X \to Y$ where $Y$ is a nonlinear function of $X$ and independent additive noise $E_Y$:

$$\begin{aligned} Y &= f_Y(X) + E_Y \\ X &\sim P_X, \ E_Y \sim P_{E_Y}, \end{aligned} \quad (7)$$

where $X$ and $E_Y$ are independent. Then, the density $P_{X,Y}$ is said to be induced by the additive noise model (ANM) from $X$ to $Y$ (Mooij et al. 2016). The alternative additive noise model between $X$ and $Y$ is the additive noise model $Y \to X$:

$$\begin{aligned} X &= f_X(Y) + E_X \\ Y &\sim P_Y, E_X \sim P_{E_X}, \end{aligned} \quad (8)$$

where $Y$ and $E_X$ are independent.

If the density $P_{X,Y}$ is induced by the ANM $X \to Y$, but not by the ANM $Y \to X$, then the ANM $X \to Y$ is identifiable. Independence of cause and mechanism states that the conditional distribution $P_{Y|X}$ contains no information about the distribution of causal $P_X$. In other words, $P_X$ and $P_{Y|X}$ are algorithmically independent:

$$I(P_X : P_{Y|X}) \stackrel{+}{=} 0. \quad (9)$$

It is from the model (7) that $P_{Y|X} = P_{E_Y}$. Therefore, from equation (9) we obtain

$$I(P_X : P_{E_Y}) \stackrel{+}{=} 0, \tag{10}$$

which implies

$$I(x; E_Y) = 0. \tag{11}$$

Mutual information of zero between the cause $X$ and residual variable $E_Y$ shows that $X$ and $E_Y$ are independent. Therefore, algorithmic independence between the distribution of cause $X$ and conditional distribution $P_{Y|X}$ of effect given the cause is equivalent to the independence of two random variables $X$ and $E_Y$ in the ANM. Peters et al. (2017) showed that a joint distribution $P_{X,Y}$ does not admit an ANM in both directions at the same time under some quite generic conditions.

To illustrate that ANMs are generally identifiable, i.e., a joint distribution only admits an ANM in one direction, we plotted Figures 1 and 2. The data in Figures 1 and 2 were generated by $y = x^3 + e_Y$, where $e_Y$ is uniformly distributed in $[-1, 1]$.

The joint distribution satisfied an ANM $X \to Y$, but did not admit an ANM $Y \to X$. Figure 1 clearly demonstrated that conditional distribution $P_{Y|X}$ did not depend on the cause $X$. However, conditional distribution $P_{X|Y}$, in deed, depended on $Y$ (Figure 2). In other words, it violated the principal of independence of cause and mechanism. The joint distribution in this example only admitted an ANM in only one direction $X \to Y$.

Empirically, if the ANM $X \to Y$ fits the data, then we infer that $X$ causes $Y$, or if the ANM $Y \to X$ fits the data, then $Y$ causes $X$ will be concluded. Although this statement cannot be rigorously proved, in practice, this principle will provide the basis for bivariate cause discovery (Mooij et al. 2016). To implement this principal, we need to develop statistical methods for assessing whether the additive noise model fits the data or not.

Now we summarize procedures for using ANM to assess causal relationships between two variables. Two variables can be two gene expressions, or one gene expression and one methylation level of CpG site, or an imaging signal of one brain region and a functional principal score of gene. Divide the dataset into a training data set by specifying $D_{train} = \{Y_n, X_n\}, Y_n = [y_1,...,y_n]^T, X_n = [x_1,...,x_n]^T$ for fitting the model and a test data set $D_{test} = \{\widetilde{Y}_m, \widetilde{X}_m\}, \widetilde{Y}_m = [\widetilde{y}_1,...,\widetilde{y}_m]^T, \widetilde{X}_m = [\widetilde{x}_1,...,\widetilde{x}_m]^T$ for testing the independence, where $n$ is not necessarily equal to $m$.

*Algorithm for causal discovery with two continuous variables is given below.*

Step 1. Regress $Y$ on $X$ using the training dataset $D_{train}$ and non-parametric regression methods:

$$Y = \hat{f}(x) + E_Y. \tag{12}$$

Step 2. Calculate residual $\hat{E}_Y = Y - \hat{f}(x)$ using the test dataset $D_{test}$ and test whether the residual $\hat{E}_Y$ is independent of causal $X$ to assess the ANM $X \to Y$.

Step 3. Repeat the procedure to assess the ANM $Y \to X$.

Step 4. If the ANM in one direction is accepted and the ANM in the other is rejected, then the former is inferred as the causal direction.

There are many non-parametric methods that can be used to regress $Y$ on $X$ or regress $X$ on $Y$. For example, we can use smoothing spline regression methods (Wang 2011), B-spline (Wang 2017) and local polynomial regression (LOESS, see Cleveland , 1979).

Covariance can be used to measure association, but cannot be used to test independence between two variables. A covariance operator can measure the magnitude of dependence, and is a useful tool for assessing dependence between variables. Specifically, we will use the Hilbert-Schmidt norm of the cross-covariance operator or its approximation, the Hilbert-Schmidt independence criterion (HSIC) to measure the degree of dependence between the residuals and potential causal variable (Gretton et al. 2005; Mooij et al. 2016).

*Calculation of the HSIC consists of the following steps.*

Step 1: Use test data set to compute

$$y_i = \hat{f}(x_i) + E_Y(i), i = 1, \ldots, m.$$

Step 2: compute the residuals:

$$\varepsilon_i = E_Y(i) = y_i - \hat{f}(x_i), = 1, \ldots, m.$$

Step 3: Select two kernel functions $k_E(\varepsilon_i, \varepsilon_j)$ and $k_x(x_1, x_2)$. Compute the Kernel matrices:

$$K_{E_Y} = \begin{bmatrix} k_E(\varepsilon_1, \varepsilon_1) & \cdots & k_E(\varepsilon_1, \varepsilon_m) \\ \vdots & \vdots & \vdots \\ k_E(\varepsilon_m, \varepsilon_1) & \cdots & k_E(\varepsilon_m, \varepsilon_m) \end{bmatrix}, K_x = \begin{bmatrix} k_x(x_1, x_1) & \cdots & k_x(x_1, x_m) \\ \vdots & \vdots & \vdots \\ k_x(x_m, x_1) & \cdots & k_x(x_m, x_m) \end{bmatrix}.$$

Step 4: compute the HSCI for measuring dependence between the residuals and potential causal variable.

$$HSIC^2(E_Y, X) = \frac{1}{m^2} \text{Tr}(K_{E_Y} H K_X H),$$

where $H = I - \frac{1}{m}\mathbf{1}_m \mathbf{1}_m^T$, $\mathbf{1}_m = [1,1,\ldots,1]^T$ and Tr denotes the trace of the matrix.

*In summary, the general procedure for bivariate causal discovery is given as follows (Mooij et al. 2016):*

Step 1: Divide a data set into a training data set $D_{train} = \{Y_n, X_n\}$ for fitting the model and a test data set $D_{test} = \{\tilde{Y}_m, \tilde{X}_m\}$ for testing the independence.

Step 2: Use the training data set and nonparametric regression methods

    (a) Regress $Y$ on $X$: $Y = f_Y(x) + E_Y$ and

    (b) Regress $X$ on $Y$: $X = f_X(y) + E_X$.

Step 3: Use the test data set and estimated nonparametric regression model that fits the training data set $D_{train} = \{Y_n, X_n\}$ to predict residuals:

(a) $\hat{E}_{Y_X} = \tilde{Y} - \hat{f}_Y(\tilde{X})$

(b) $\hat{E}_{X_Y} = \tilde{X} - \hat{f}_X(\tilde{Y})$.

Step 4: Calculate the dependence measures $HSIC^2(E_Y, X)$ and $HSIC^2(E_X, Y)$.

Step 5: Infer causal direction:

$$X \to Y \text{ if } HSIC^2(E_Y, X) < HSIC^2(E_X, Y); \tag{13}$$

$$Y \to X \text{ if } HSIC^2(E_Y, X) > HSIC^2(E_X, Y). \tag{14}$$

If $HSIC^2(E_Y, X) = HSIC^2(E_X, Y)$, then causal direction is undecided.

We do not have closed analytical forms for the asymptotic null distribution of the HSIC and hence it is difficult to calculate the P-values of the independence tests. To overcome these limitations, the permutation/bootstrap approach can be used to calculate the P-values of the causal test statistics. The null hypothesis is

$H_0$: no causations $X \to Y$ and $Y \to X$ (Both $X$ and $E_Y$ are dependent, and $Y$ and $E_X$ are dependent).

Calculate the test statistic:

$$T_C = |HSIC^2(E_Y, X) - HSIC^2(E_X, Y)|. \tag{15}$$

Assume that the total number of permutations is $n_p$. For each permutation, we fix $x_i, i = 1, \ldots, m$ and randomly permutate $y_i, i = 1, \ldots, m$. Then, fit the ANMs and calculate the residuals $E_X(i), E_Y(i), i = 1, \ldots, m$ and test statistic $T_C$. Repeat $n_p$ times. The P-values are defined as the proportions of the statistic $\tilde{T}_C$ (computed on the permuted data) greater than or equal to $\hat{T}_C$ (computed on the original data $D_{TE}$). After causation is identified, we then use equations (13) and (14) to infer causal directions $X \to Y$ or $Y \to X$.

**Correlation and Causation**

In everyday language, correlation and association are used interchangeably. However, correlation and association are different terminologies. Correlation is to characterize the trend pattern between two variables, particularly; the Pearson correlation coefficient measures linear trends while association characterizes the simultaneous occurrence of two variables. In this paper, association is equivalent to linear correlation. We investigate the relationships between causation and association. The association between two continuous variables is defined as a linear regression model:

$$Y = \beta X + \varepsilon, \tag{16}$$

where $\beta \neq 0$.

The causation $X \to Y$ is identified by the ANM:

$$Y = f(X) + \varepsilon, \ X \perp\!\!\!\perp \varepsilon. \tag{17}$$

In classical statistics, if we assume that both variables $X$ and $\varepsilon$ follow a normal distribution, then $cov(X, \varepsilon) = 0$ if and only if $X$ and $\varepsilon$ are independent. If $X$ and $\varepsilon$ are not normal variables, this statement will not hold. For general distribution, we extend the concept of covariance to cross covariance operator $\tilde{C}_{X\varepsilon}$ (Zhang et al. 2017). It is shown that for the general distributions of $X$ and $\varepsilon$, $\tilde{C}_{X\varepsilon} = 0$ if and only if $X$ and $Y$ are independent (Mooij et al. 2016).

Let $h$ and $g$ be any two nonlinear functions. $\tilde{C}_{X\varepsilon} = 0$ is equivalent to (Gretton et al. 2005)

$$\max cov(h(X), g(\varepsilon)) = cov\left(h(X), g(Y - f(X))\right) = 0, \quad (18)$$

Subject to $||h|| = 1, ||g|| = 1$.

Now we give samples to illustrate existence of three cases: a) both correlation and causation $X \to Y$, b) causation $X \to Y$, but no correlation and c) correlation, but no causation $X \to Y$.

*a) Both correlation and causation $X \to Y$.*

We consider a special case: $Y = f(X)$. When $Y = f(X)$, equation (18) holds, which implies $X \to Y$. If we assume that $h(X) = X$ and $g(Y - f(X)) = Y - f(X)$, then equation (18) holds and implies that

$$cov(X, Y) = cov(X, f(X)). \quad (19)$$

If we further assume $f(X) = \beta X$, then equation (19) implies

$$\beta = \frac{cov(X,Y)}{Var(X)}. \quad (20)$$

This is estimation of linear regression coefficient.

*b) Causation $X \to Y$, but no correlation*

Consider the model:

$$Y = 5X^2 + \varepsilon,$$

where $X$ follows a uniform distribution between $-2$ and $2$ and $\varepsilon$ follows a uniform distribution between $-1$ and $1$.

Figure 9 plotted functions $Y = 5X^2 + \varepsilon$. Assume that 2,000 subjects were sampled. Permutation was used to calculate P-value for testing causation. We found that the Pearson correlation was $-0.00070$ and P-value for testing causation $X \to Y$ was $10^{-5}$. This example showed the presence of causation, but lack of association (correlation near zero).

c) *Correlation, but no causation $X \to Y$.*

Assume that correlation $\rho_{XY} = \frac{Cov(X,Y)}{\sqrt{Var(X)Var(Y)}}$ exists. Take $h(X) = X^2$ and $g(Y - f(X)) = (Y - f(X))^2$. Then, $Cov(h(X), g(Y - f(X))) > 0$.

Thus, $\max cov(h(X), g(\varepsilon)) = cov\left(h(X), g(Y - f(X))\right) > 0$.

Equation (18) does not hold. There is no causation in this scenario.

Similar conclusions hold for $Y \to X$.

**Simulations**

To investigate their feasibility for causal inference, the ANMs were applied to simulation data. Similar to Nowzohour and Bühlmann (2016), we considered three nonlinear functions: quadratic, exponential and logarithm functions and two random noise variables: normal and $t$ distribution. We assumed that the cause $X$ follows a normal distribution $N(0,1)$.

First we consider two models with a quadratic function and two types of random noise variables, normal $N(0,1)$ and $t$ distribution with 5 degrees of freedom:

Model 1:

$$Y = X + b \cdot x^2 + \epsilon_1,$$

where the parameter $b$ ranges from -10 to 10 and $\varepsilon_1$ is distributed as $N(0,1)$.

Model 2:

$$Y = X + b \cdot x^2 + \epsilon_2$$

where the parameter $b$ is defined as before and $\varepsilon_2$ is distributed as $t$ distribution with 5 degrees of freedom.

The parameter space $b \in [-10, 10]$ was discretized. For each grid point, 1,000 simulations were repeated. For each simulation, 500 samples were generated. The ANMs were applied to the generated data. Smoothing spline is used to fit the functional model. The true causal direction is the forward model: $X \to Y$. The false decision rate was defined as the proportion of times when the backwards model $Y \to X$ is wrongly chosen by the ANMs. Figures 3 and 4 presented false decision rate as a function of the parameter $b$ for the models 1 and 2, respectively. We observed from Figures 3 and 4 that the false decision rate reached its maximum 0.5 when $b = 0$. This showed that when the model is close to linear, the ANMs could not identify the true causal direction. However, when $b$ moved away from 0, the false decision rates approached 0 quickly. This showed that when the data were generated by nonlinear models, with high probability, we can accurately identify the true causal directions.

To further confirm these observations, we consider another two nonlinear functions.

Model 3:

$$Y = X + b \log(|X|) + \varepsilon_1,$$

Model 4:

$$Y = X + b \log(|X|) + \varepsilon_2,$$

Model 5:

$$Y = X + b \cdot e^X + \varepsilon_1,$$

Model 6:

$$Y = X + b \cdot e^X + \varepsilon_2,$$

where the parameter $b$ and the noise variables $\varepsilon_1$ and $\varepsilon_2$ were defined previously.

The false decision rates of the ANMs for detecting the true causal direction $X \to Y$ for the models 3, 4, 5 and 6 were presented in Figures 5, 6 7 and 8, respectively. Again, the observations for the models 1 and 2 still held for the models 3, 4, 5 and 6. When the data were generated by nonlinear models, we can accurately identify the true causal directions. However, when the data were generated by linear models, the false decision rates reached 0.5, which was equivalent to random guess.

**Real Data Analysis**

Regulation of gene expression is a complex biological process. Large-scale regulatory network inference provides a general framework for comprehensively learning regulatory interactions, understanding the biological activity, devising effective therapeutics, identifying drug targets of complex diseases and discovering the novel pathways. Uncovering and modeling gene regulatory networks are one of the long-standing challenges in genomics and computational biology. Various statistical methods and computational algorithms for network inference have been developed. The ANMs can also be applied to inferring gene regulatory networks using gene expression data. Similar to co-gene expression networks where correlations are often used to measure dependence between two gene expressions, the ANMs can be used to infer regulation direction, i.e., whether changes in expression of gene $X$ causes changes in expression of gene $Y$ or vise verse changes in expression of gene $Y$ causes changes in expression of gene $X$.

The ANMs were applied to Wnt signaling pathway with RNA-Seq of 79 genes measured in 447 tissue samples. For comparisons, the SEMs integrating with integer programming (Xiong 2018), causal additive model (CAM) (Bühlmann et al. 2014), PC algorithm (Tang et al. 2011) , random network,  glasso (Friedman et al. 2015),  and Weighted Correlation Network Analysis (WGCNA) (Langfelder and Horvath, 2008) were also included in the analysis. The results were summarized in Table 1. True directed path was defined as the paths that matched KEGG paths with directions. True undirected path was defined as the paths that matched KEGG paths with or without directions. Detection accuracy was defined as the proportion of the number of true paths detected over the number of all paths detected.

Figure 10 presented the ANM-inferred network structure of the Wnt pathway. The green lines represented the inferred paths consistent to the KEGG while the gray ones represented the inferred edges absent in the KEGG. The ANM, CAM, SEM, PC, and random network methods inferred directed networks, and Glasso and WGCNA association methods inferred undirected networks. We took the structure of Wnt in the KEGG as the true structure of the Wnt in nature. We observed from Table 1 that the ANM more accurately inferred the network structure of the Wnt than the other six statistical and computational methods for identifying directed or undirected networks. Table 1 also showed that the accuracy of widely used Glasso and WGCNA algorithms for identifying the structure of Wnt was even lower than that of random networks, however, the accuracy of the ANM was much higher than that of random networks.

To evaluate their performance for causal inference, the ANMs were applied to the Alzheimer's Disease Neuroimaging Initiative (ADNI) data with 91 individuals with Diffusion Tensor Imaging (DTI) and cholesterol phenotypes measured at four time points: baseline, 6 months, 12 months and 24 months. After normalization and image registration, the dimension of a single

DTI image is $91 \times 109 \times 91$. Three dimensional functional principal component analysis (3D-FPC) was used to summarize imaging signals in the brain region (Lin et al. 2015), because of the technical difficulty and operational cost, only 44 of the 91 individuals have all the DTI imaging data at all the four data points. Based on our own analysis experience, usually the first one or two 3D-FPC scores can explain more that 95% of the variation of the imaging signals in the region. To evaluate the performance of 3D-FPC for imaging signal feature extraction, we present Figures 11A and 11B. Figure 11A is a layer of the FA map of the DTI image from a single individual and the dimension of this image is $91 \times 109$. A total of 91 images were used to calculate the 3D-FPC scores. Figure 11B was the reconstruction of the same layer of the FA map of the DTI image from the same individual in Figure 11A using 5 FPC scores. Comparing Figure 11A with Figure 11B, we can see that these two images are very similar indicating that the 3D-FPC score is an effective tool to represent the image features.

To investigate feasibility of image imputation by using a mixed strategy of 3D-FPC scores and matrix completion, we used the DTI image of the 44 individuals who have measurement at all four time points as the investigation dataset. Since at baseline, the DTI image of all individuals was available, we did not have missing value problems. We only need to impute images at 6, 12 and 24 months for some individuals. We randomly sampled 20 individuals assuming that their imaging data were missing. Matrix completion methods were used to impute missing images (Thung et al. 2018). To perform 3D FPCA, all missing imaging signals at 6, 12 and 24 months of the individuals were replaced by their imaging signals at the baseline. Then, 3D FPCA was performed on the original images and replaced images of 44 individuals at all time points (base line, 6, 12 and 24 months). The FPC scores of 22 individuals without missing images were used for matrix completion. The imputed FPC score were then used to form reconstruction of the DTI images. To evaluate performance of the above image imputation, we presented Figure 12 that was the reconstruction of the DTI image in Figure 11A. We observed from these figures that the imputed image captured the majority of the information in the original DTI image data.

After image imputation, DTI images at all four points and cholesterol and working memory of 91 individuals were available. The DTI images were segmented into 19 brain regions using the Super-voxel method (Achanta et al. 2012). Three dimensional functional principal component analysis was used to summarize imaging signals in the brain region (Lin et al. 2015). The ANMs were used to infer causal relationships between cholesterol, or working memory and image where only first FPC score (accounting for more than 95% of the imaging signal variation in the segmented region) was used to present the imaging signals in the segmented region. Table 2 presented P-values for testing causation (cholesterol $\rightarrow$ image variation) and association of cholesterol with images of 19 brain regions where the canonical correlation method was used to test association (Lin et al. 2017). Two remarkable features emerged. First, we observed both causation and association of cholesterol with imaging signal variation at 24 months in the temporal L hippocampus (P-value for causation < 0.00013, P-value for association < 0.00007) and temporal R hippocampus regions (P-value for causation < 0.0165, P-value for association < 0.0044), and only association of cholesterol with imaging signal variation at 12 months in the temporal L region (P-value for causation < 0.5262, P-value for association < 0.0038). Figures 13A and 13B presented the curves of cholesterol level of an AD patient and average cholesterol level of normal individuals, and images at baseline, 6 months, 12 months and 24 months of the temporal L hippocampus of an individual with AD diagnosed at 24 months time point, respectively. Figures 14A and 14C presented the curves of cholesterol level of an individual with AD diagnosed at 24 months' time point and average cholesterol levels of normal individuals, and images at baseline, 6 months, 12 months and 24 months of the Temporal R regions of an individual with AD diagnosed at 24 months' time point, respectively. Figures 13 and 14 showed that images of the temporal L hippocampus and Temporal R regions at 24 months became black, which indicated that temporal L hippocampus and temporal R regions were damaged by the high cholesterol. Second, we observed only association of cholesterol with imaging signal variation at 12 and 24 months in the Occipital_Mid brain region (P-value < 0.0003 at 12 months, P-value < 0.00004 at 24 months), but no causation (P-value < 0.6794 at 12 months, P-value < 0.1922 at 24 months). Figure 15 showed images of the occipital lobe region. We observed that there were no significant imaging signal variation in the occipital lobe region. This strongly demonstrates that association may not provide information on unravelling mechanism of complex phenotypes.

In our phenotype-image studies, we also identified causal relationships between working memory and activities of the temporal R (hippocampus) at 24 months with P-value < 0.00014) (image → working memory), but identified no association of working memory with imaginal signal variation in the temporal R (hippocampus) region (P-value < 0.5904) (Table 3). Figure 14C showed the weak imaging signal or decreased neural activities in the temporal R (hippocampus) region at 24 months and Figure 14B showed lower working memory measure of an AD patient than the average working memory measurements of normal individuals at 24 months. This demonstrated that the decreased neural activities in the temporal R (hippocampus) region deteriorated working memory of the AD patient. This result provided evidence that causation may be identified in the absence of association signals. These observations can be confirmed from the literature. It was reported that cholesterol level impacted the brain white matter connectivity in the temporal gyrus (Haltia et al. 2007) and was related to AD (Sjogren et al. 2005; Teipel et al. 2006). Abnormality in working memory was observed in patients with temporal lobe epilepsy (Stretton et al. 2013).

**Discussion**

The major purpose of this paper is to address several issues for shifting the paradigm of genetic analysis from association analysis to causal inference and to focus on causal discovery between two variables. The first issue is the basic principles for causal inference from observational data only. Typical methods for unravelling cause and effect relationships are interventions and controlled experiments. Unfortunately, the experiments in human genetics are unethical and technically impossible. In the past decade, the new principles for causal inference from pure observational data have been developed. The philosophical causal principle assumes that nature consists of autonomous and independent causal generating process modules and attempts to replace causal faithfulness by the assumption of Independence of Cause and Mechanism (ICM). In other words, causal generating processes of a system's variables are independent. If we consider two variables. The ICM states that distribution of cause and conditional distribution of effect, given that cause are independent.

The second issue is how to measure independence (or dependence) between two distributions. Statistics only provides tools for measuring independence between two random variables. There are no measures or statistics to test independence between two distributions. Therefore, we introduce algorithmic information theory that can offer notion and mathematical formulation of independence between two distributions or independence of mechanisms. We use algorithmic mutual information to measure independence between two distributions which can be used to assess causal relationships between two variables. Algorithmically independent conditional implies that the joint distribution has a shorter description in causal direction than in non-causal direction.

The third issue is to develop causal models that can easily assess algorithmic independent conditions. The algorithmic independent condition states that the direction with the lowest Kolmogorov complexity can be identified to be the most likely causal direction between two random variables. However, it is well known that the Kolmogorov complexity is not computable (Budhathoki and Vreeken, 2017). Although stochastic complexity was proposed to approximate Kolmogorov complexity via the Minimum Description Length (MDL) principle, it still needs heavy computations. The ANM was developed as practical causal inference methods to implement algorithmically independent conditions. We showed that algorithmic independence between the distribution of cause $X$ and conditional distribution $P_{Y|X}$ of effect given the cause is equivalent to the independence of two random variables $X$ and $E_Y$ in the ANM.

The fourth issue is the development of test statistics for bivariate causal discovery. The current ANM helps to break the symmetry between two variables $X$ and $Y$. Its test statistics are designed to identify causal directions: $X \rightarrow Y$ or $Y \rightarrow X$. Statistics and methods for calculation of P-values for testing the causation between two variables have not been developed. To address this issue, we have developed a new statistic to directly test for causation between two variables and a permutation method for the calculation of P-value of the test.

The fifth issue is the power of the ANM. The challenge arising from bivariate causal discovery is whether the ANM has enough power to detect causation between two variables. To investigate their feasibility for causal inference, the ANMs were applied to simulation data. We considered three nonlinear functions: quadratic, exponential and logarithm functions and two random noise variables: normal and t distribution. We showed that the ANM had reasonable power to detect existence of causation between two variables. To further evaluate its performance, the ANM was also applied to reconstruction of the Wnt pathway using gene expression data. The results demonstrated that the ANM had higher power to infer gene regulatory networks than six other statistical methods using KEGG pathway database as gold standard.

The sixth issue is how to distinguish association from causation. In everyday language, correlation and association are used interchangeably. However, correlation and association are different terminologies. Correlation is to characterize the trend pattern between two variables, particularly; the Pearson correlation coefficient measures linear trends, while association characterizes the simultaneous occurrence of two variables. The widely used notion of association often indicates the linear correlation. When two variables are linearly correlated we say that there is association between them. Pearson correlation or its equivalent, linear regression is often used to assess association. Causation between two variables is defined as independence between the distribution of cause and conditional distribution of the effect, given cause. In the nonlinear ANM, the causal relation is assessed by testing independence between the cause variable and residual variable in the nonlinear ANM. We investigated the relationships between causation and association (linear correlation). Some theoretical analysis and real trait-imaging data analysis showed that there were three scenarios: (1) presence of both association and causation between two variables, (2) presence of association, while absence of causation and (3) presence of causation, while lack of association in causal analysis.

Finally, in real imaging data analysis, we showed that causal traits changes the imaging signal variation in the brain regions. However, the traits that were associated with the imaging signal in the brain regions did not change imaging signals in the region at all.

The experiences in association analysis in the past several decades strongly demonstrate that association analysis is lack of power to discover the mechanisms of the diseases and provide powerful tools for medicine. It is time to shift the current paradigm of genetic analysis from shallow association analysis to more profound causal inference. Transition of analysis from association to causation raises great challenges. The results in this paper are considered preliminary. A large proportion of geneticists and epidemiologists have serious doubt about the feasibility of causal inference in genomic and epigenomic research. Causal genetic analysis is in its infantry. The novel concepts and methods for causal analysis in genomics, epigenomics and imaging data analysis should be developed in the genetic community. Large scale simulations and real data analysis for causal inference should be performed. We hope that our results will greatly increase the confidence in genetic causal analysis and stimulate discussion about whether the paradigm of genetic analysis should be changed from association to causation or not.

**Acknowledgments**

The project described was partially supported by Grant from National Natural Science Foundation of China (No. 81373100).

**Author Contribution Statement**

Rong Jiao (RJ) : Perform data analysis and write paper

Nan Lin (NL): Perform data analysis

Zixin Hu (ZH): Perform data analysis

David A Bennett (DAB): Provide Data and Result Interpretation

Li Jin (LJ) : Design project

Momiao Xiong (MX) : Design project and write paper

# References

Mooij, J.M., Peters, J., Janzing, D., Zscheischler, J., & Schölkopf, B. (2016). Distinguishing Cause from Effect Using Observational Data: Methods and Benchmarks. *Journal of Machine Learning Research. 17*, 32:1-32:102.

Pearl, J. (2018). Theoretical impediments to machine learning with seven sparks from the causal revolution. *arXiv preprint arXiv:1801.04016*.

Glymour, C. (2015). The causal revolution, observational science and big data. *Lecture presented at Ohio University in the History and Philosophy of Science series, Athens, Ohio.*

Zhang, K., Schölkopf, B., Spirtes, P., Glymour, C. (2018). Learning causality and causality-related learning. *National Science Review*. 5. 26-29. doi:10.1093/nsr/nwx137.

Le, T., Hoang, T., Li, J., Liu, L., Liu, H. & Hu, S. (2016). A fast PC algorithm for high dimensional causal discovery with multi-core PCs. *IEEE/ACM Transactions on Computational Biology and Bioinformatics*. doi:10.1109/TCBB.2016.2591526.

Nowzohour, C. and Bühlmann, P. (2016). Score-based causal learning in additive noise models. *Statistics*. 50, 471-485.

Peters J, Mooij J, Janzing D, Schoelkopf B. (2011). Identifiability of Causal Graphs using Functional Models. In *Proceedings of the 27th Annual Conference on Uncertainty in Artificial Intelligence (UAI)*.

Peters, J. & Bühlman, P. (2014). Identifiability of Gaussian Structural Equation Models with Equal Error Variances. *Biometrika*. 101, 219-228.

Shimizu, S., Hoyer, P. O., Hyvärinen, A., & Kerminen, A. (2006). A linear non-gaussian acyclic model for causal discovery. *Journal of Machine Learning Research.* 7, 2003-2030.

Hoyer, P. O., Janzing, D., Mooij, J. M., Peters, J., & Schölkopf, B. (2009). Nonlinear causal discovery with additive noise models. *Advances in neural information processing systems*. 689-696.

Peters, J., Mooij, J. M., Janzing, D., & Schölkopf, B. (2014). Causal discovery with continuous additive noise models. *The Journal of Machine Learning Research*. 15, 2009-2053.

Zhang, K., Hyvärinen, A. (2009). On the identifiability of the post-nonlinear causal model. In *Proceedings of the 25th Annual Conference on Uncertainty in Artificial Intelligence (UAI)*.

Peters, J., Janzing, D., Schölkopf, B. (2017). Elements of Causal Inference - Foundations and Learning Algorithms Adaptive Computation and Machine Learning Series. Cambridge, MA: The MIT Press.

Besserve, M., Shajarisales, N., Schölkopf, B., Janzing, D. (2017). Group invariance principles for causal generative models. *arXiv preprint arXiv:1705.02212.*

Schölkopf, B. and Janzing, D. and Peters, J. and Sgouritsa, E. and Zhang, K. and Mooij, J. (2012). On Causal and Anticausal Learning. *Proceedings of the 29th International Conference on Machine Learning.* 1255-1262.

Lemeire J and Janzing D. (2013). Replacing causal faithfulness with algorithmic independence of conditionals. *Minds and Machines*. 23(2):227–249.

Shajarisales, N., Janzing, D., Schölkopf, B., Besserve, M. (2015). Telling cause from effect in deterministic linear dynamical systems. *Proceedings of the 32nd International Conference on Machine Learning*. 37, 285–294.

Janzing, D. and Sch¨olkopf, B. (2010). Causal inference using the algorithmic Markov condition. *IEEE Transactions on Information Theory*. 56, 5168–5194.

Parascandolo, G., Rojas-Carulla, M., Kilbertus, N., Schölkopf, B. (2017). Learning Independent Causal Mechanisms. *arXiv preprint arXiv:1712.00961.*

Kolmogorov, A. (1965). Three approaches to the quantitative definition of information. *Problems of Information Transmission*. 1, 3–11.

Grunwald, P. D., & Vitanyi, P. M. B. (2004). Shannon Information and Kolmogorov complexity. *IEEE Trans. Information Theory, arXiv:cs/0410002.*

Wang YD. (2011). Smoothing splines: methods and applications. New York: CRC Press.

Wang, W., Yan, J. (2017). splines2: Regression Spline Functions and Classes. *R package version 0.2.7.*

Cleveland WS. (2012). Robust locally weighted regression and smoothing scatterplots. *Journal of the American Statistical Association.* 74: 368, 829-836.

Gretton, A., Bousquet, O., Smola, A., and Schölkopf B. (2005). Measuring statistical dependence with Hilbert–Schmidt norms. In *Proceedings of the International Conference on Algorithmic Learning Theory*, 63–77.

Nowzohour, C. and Bühlmann, P. (2016). Score-based causal learning in additive noise models. *Statistics*. 50: 471–485.

Achanta R, Shaji A, Smith K, Lucchi A, Fua P, Süsstrunk S. (2012). SLIC Superpixels Compared to State-of-the-Art Superpixel Methods. *IEEE Transactions on Pattern Analysis and Machine Intelligence.* 34: 2274-82.

Lin N, Jiang J, Guo S, Xiong M. (2015). Functional Principal Component Analysis and Randomized Sparse Clustering Algorithm for Medical Image Analysis. *PLOS ONE*. 10(7): e0132945.

Thung KH, Yap PT, Adeli E, Lee SW, Shen D; Alzheimer's Disease Neuroimaging Initiative. (2018). Conversion and time-to-conversion predictions of mild cognitive impairment using low-rank affinity pursuit denoising and matrix completion. *Med Image Anal*. 45, 68-82.

Lin N, Zhu Y, Fan R and Xiong MM. (2017). A Quadratically Regularized Functional Canonical Correlation Analysis for Identifying the Global Structure of Pleiotropy with NGS Data. *PLOS Computational Biology*. 13(10): e1005788.

Zhang, Q., Filippi, S., Gretton, A., and Sejdinovic, D. (2017). Large-Scale Kernel Methods for Independence Testing. *Statistics and Computing*, pp.1-18.

Langfelder P, Horvath S. (2008). WGCNA: an R package for weighted correlation network analysis. *BMC Bioinformatics*. 9, 559.

Friedman, J., Hastie, T., Tibshirani, R. (2008). Sparse inverse covariance estimation with the graphical lasso. *Biostatistics*. 9: 432–441.

Bühlmann, P. Peters, J., Ernest J. (2014). CAM: Causal Additive Models, high-dimensional order search and penalized regression. *Annals of Statistics*. 42, 2526-2556.

Xiong MM. (2018). Big data in omics and image: integrated analysis and causal inference. CRC Press.

Tan M, Alshalalfa M, Alhajj R, Polat F. (2011). Influence of prior knowledge in constraint-based learning of gene regulatory networks. *IEEE/ACM Trans Comput Biol Bioinform*. 8: 130-42.

Haltia LT, Viljanen A, Parkkola R, Kemppainen N, Rinne JO, Nuutila P, et al. (2007). Brain White Matter Expansion in Human Obesity and the Recovering Effect of Dieting. *The Journal of Clinical Endocrinology & Metabolism*. 92(8): 3278-84. doi: 10.1210/jc.2006-2495.

Sjogren M, Blennow K. (2005). The link between cholesterol and Alzheimer's disease. *World J Biol Psychiatry*. 6(2): 85-97.

Teipel SJ, Pruessner JC, Faltraco F, Born C, Rocha-Unold M, Evans A, et al. (2006). Comprehensive dissection of the medial temporal lobe in AD: measurement of hippocampus, amygdala, entorhinal, perirhinal and parahippocampal cortices using MRI. *J Neurol*. 253(6): 794-800. doi: 10.1007/s00415-006-0120-4.

Stretton J, Winston GP, Sidhu M, Bonelli S, Centeno M, Vollmar C, et al. (2013). Disrupted segregation of working memory networks in temporal lobe epilepsy. *Neuroimage Clin*. 2: 273-81. doi: 10.1016/j.nicl.2013.01.009.

**Tables**

Table 1. Accuracy of the ANMs and other six methods for inferring Wnt pathway.

| Methods | Detection Accuracy | |
|---|---|---|
| | Directed Paths | Paths with or without directions |
| Pairwise ANM | 38% | 46% |
| CAM | 16% | 24% |
| SEM | 20% | 26% |
| PC Algorithm | 21.57% | 39.22% |
| Random Network | 25.41% | 30.64% |
| Glasso | | 28% |
| WGCNA association | | 22% |

Table 2. P-values for assessing association and causal relationships between the cholesterol and brain region.

| | Baseline | | 6 Months | | 12 Months | | 24 Months | |
|---|---|---|---|---|---|---|---|---|
| | Causal | Association | Causal | Association | Causal | Association | Causal | Association |
| Frontal_Inf_R | 0.5699 | 0.4318 | 0.2927 | 0.9390 | 0.2169 | 0.7145 | 0.6624 | 0.1580 |
| Frontal_Sup_Mid_L | 0.4061 | 0.5539 | 0.0203 | 0.0301 | 0.6905 | 0.8670 | 0.3316 | 0.9664 |
| Insula_L | 0.9274 | 0.4602 | 0.2766 | 0.3102 | 0.5396 | 0.2724 | 0.7734 | 0.6819 |
| Fusiform_L | 0.3253 | 0.6601 | 0.8358 | 0.1778 | 0.5720 | 0.6238 | 0.8411 | 0.4510 |
| Insula_R | 0.3853 | 0.2367 | 0.6093 | 0.8874 | 0.0109 | 0.1218 | 0.2575 | 0.1832 |
| Temporal_R | 0.3740 | 0.7487 | 0.2997 | 0.3214 | 0.2813 | 0.8856 | 0.0165 | 0.0044 |
| Occipital_Mid | 0.7275 | 0.3344 | 0.8082 | 0.4159 | 0.6794 | 0.0003 | 0.1922 | 0.00004 |
| Temporal_L | 0.1455 | 0.4873 | 0.5384 | 0.9752 | 0.5262 | 0.0038 | 0.0001 | 0.0001 |
| Frontal_L_R | 0.1673 | 0.9822 | 0.8928 | 0.9269 | 0.3784 | 0.4762 | 0.5832 | 0.8093 |
| Frontal & Temp_L | 0.6067 | 0.4698 | 0.9643 | 0.3847 | 0.2945 | 0.9249 | 0.5057 | 0.1937 |
| Lingual | 0.2625 | 0.5307 | 0.8354 | 0.0834 | 0.7238 | 0.8036 | 0.2230 | 0.5510 |
| Cingulum | 0.6232 | 0.6483 | 0.3061 | 0.1381 | 0.0587 | 0.7611 | 0.3581 | 0.6024 |
| Precentral_R | 0.7113 | 0.4946 | 0.7263 | 0.0948 | 0.1565 | 0.6969 | 0.5169 | 0.6388 |
| Frontal_Inf_L | 0.9167 | 0.9260 | 0.5886 | 0.0138 | 0.3091 | 0.0929 | 0.3568 | 0.7203 |
| Occipital | 0.2444 | 0.3753 | 0.0782 | 0.9927 | 0.8490 | 0.2909 | 0.7388 | 0.4617 |
| Precuneus | 0.8480 | 0.2492 | 0.4183 | 0.9418 | 0.7208 | 0.5096 | 0.9071 | 0.8899 |
| SMP | 0.9866 | 0.1630 | 0.4416 | 0.6642 | 0.1175 | 0.3797 | 0.9788 | 0.3388 |
| Precentral_L | 0.6825 | 0.7937 | 0.4142 | 0.0759 | 0.9402 | 0.5150 | 0.5254 | 0.9770 |
| Precentral_R | 0.0488 | 0.4103 | 0.9759 | 0.9831 | 0.7251 | 0.9000 | 0.5008 | 0.0105 |

Table 3. P-values for assessing association and causal relationships between the working memory and brain region.

| | Baseline | | 6 Months | | 12 Months | | 24 Months | |
|---|---|---|---|---|---|---|---|---|
| | Causal | Association | Causal | Association | Causal | Association | Causal | Association |
| Frontal_Inf_R | 0.7515 | 0.6348 | 0.4857 | 0.5088 | 0.3709 | 0.5807 | 0.5028 | 0.0572 |
| Frontal_Sup_Mid_L | 0.2022 | 0.2877 | 0.0187 | 0.8929 | 0.2355 | 0.8327 | 0.4114 | 0.7976 |
| Insula_L | 0.0300 | 0.5539 | 0.4928 | 0.1057 | 0.8959 | 0.5846 | 0.6212 | 0.0332 |
| Fusiform_L | 0.3244 | 0.5135 | 0.0931 | 0.0503 | 0.0617 | 0.9162 | 0.6927 | 0.0741 |
| Insula_R | 0.2212 | 0.9885 | 0.7729 | 0.6777 | 0.5171 | 0.1434 | 0.7416 | 0.4923 |
| Temporal_R | 0.9042 | 0.5224 | 0.9641 | 0.6987 | 0.2813 | 0.0939 | 0.0001 | 0.5904 |
| Occipital_Mid | 0.8350 | 0.4884 | 0.0309 | 0.7277 | 0.6280 | 0.9993 | 0.2067 | 0.4716 |
| Temporal_L | 0.9491 | 0.8716 | 0.1052 | 0.4597 | 0.0001 | 0.0006 | 0.0001 | 0.5836 |
| Frontal_L_R | 0.8957 | 0.0212 | 0.2522 | 0.5165 | 0.2658 | 0.7134 | 0.1474 | 0.1720 |
| Frontal & Temp_L | 0.9189 | 0.3919 | 0.7792 | 0.1148 | 0.3951 | 0.3585 | 0.7691 | 0.7355 |
| Lingual | 0.4241 | 0.3219 | 0.4952 | 0.5941 | 0.1707 | 0.8981 | 0.8382 | 0.6736 |
| Cingulum | 0.5063 | 0.5778 | 0.0383 | 0.9534 | 0.5947 | 0.3123 | 0.1482 | 0.6307 |
| Precentral_R | 0.1398 | 0.2945 | 0.9875 | 0.5693 | 0.3247 | 0.7966 | 0.7323 | 0.7358 |
| Frontal_Inf_L | 0.8985 | 0.0989 | 0.2982 | 0.3727 | 0.8644 | 0.0363 | 0.9291 | 0.9581 |
| Occipital | 0.3828 | 0.8736 | 0.5267 | 0.8378 | 0.4624 | 0.1352 | 0.6937 | 0.1991 |
| Precuneus | 0.7215 | 0.8909 | 0.1169 | 0.5417 | 0.0406 | 0.6599 | 0.0429 | 0.9704 |
| SMP | 0.0900 | 0.7818 | 0.9407 | 0.6380 | 0.4428 | 0.3417 | 0.3151 | 0.8178 |
| Precentral_L | 0.9660 | 0.7217 | 0.6289 | 0.6630 | 0.8759 | 0.5526 | 0.8848 | 0.1713 |
| Precentral_R | 0.4051 | 0.3829 | 0.4783 | 0.5286 | 0.6365 | 0.0569 | 0.9260 | 0.5996 |

**Figure Legends**

**Figure 1** An example of joint distribution $p(x, y)$ generated by $y \coloneqq f(x) + e_Y$, where $f(x) = x^3$ and $e_Y$ is uniformly distributed in $[-1, 1]$ We perform a nonlinear regression in the directions $X \to Y$.

**Figure 2** An example of joint distribution $p(x, y)$ generated by $y \coloneqq f(x) + e_Y$, where $f(x) = x^3$ and $e_Y$ is uniformly distributed in $[-1, 1]$ We perform a nonlinear regression in the directions $Y \to X$.

**Figure 3** false decision rates as a function of the parameter $b$ for the model 1.

**Figure 4** false decision rates as a function of the parameter $b$ for the model 2.

**Figure 5** The false decision rates of the ANMs for detecting the true causal direction $X \to Y$ for the model 3.

**Figure 6** The false decision rates of the ANMs for detecting the true causal direction $X \to Y$ for the model 4.

**Figure 7** The false decision rates of the ANMs for detecting the true causal direction $X \to Y$ for the model 5.

**Figure 8** The false decision rates of the ANMs for detecting the true causal direction $X \to Y$ for the model 6.

**Figure 9** The data generated by $Y = 5X^2 + \varepsilon$, where $X$ follows a uniform distribution between $-2$ and $2$ and $\varepsilon$ follows a uniform distribution between $-1$ and $1$.

**Figure 10** The ANM-inferred network structure of the Wnt pathway. The green lines represented the inferred paths consistent to the KEGG while the gray ones represented the inferred edges absent in the KEGG.

**Figure 11(A)** A slice of the FA map from a single individual's DTI data.

**Figure 11(B)** FA map reconstruction with the first two 3D-FPC scores.

**Figure 12** Imputed FA map in Figure 11A using 3D-FPC scores and matrix completion.

**Figure 13(A)** AD and normal individuals' CHL curves.

**Figure 13(B)** Images of temporal L hippocampus region.

**Figure 14(A)** AD and normal individuals' CHL curves.

**Figure 14(B)** AD and normal individuals' working memory.

**Figure 14(C)** Images of temporal R hippocampus region.

**Figure 15(A)** AD and normal individuals' CHL curves.

**Figure 15(B)** Images of Occipital Lobe Region.

**Supplementary Note**

We show that for every joint distribution $P_{X,Y}$ of real-valued variables $X$ and $Y$, there is a nonlinear model:

$$Y = f_Y(X, N_Y), X \perp\!\!\!\perp N_Y,$$

where $f_Y$ are functions and $N_Y$ is a real-valued noise variables.

Proof.

Define the conditional cumulative distribution function:

$$F_{Y|X}(y) = P(Y \leq y | X = x) \tag{A1}$$

and let

$$N_Y = F_{Y|X}(y). \tag{A2}$$

Define its inverse function

$$F_{Y|x}^{-1}(n_Y) := \inf\{z \in R: F_{Y|x}(z) \geq n_Y\}. \tag{A3}$$

Define function

$$f_Y(x, n_Y) := F_{Y|x}^{-1}(n_Y). \tag{A4}$$

Now we make changes of variables:

$$x = x \tag{A5}$$

and

$$y = f_Y(x, n_Y). \tag{A6}$$

The Jacobian matrix of the transformation is given by

$$J = \begin{vmatrix} 1 & 0 \\ 0 & \frac{1}{F'_{Y|x}(y)} \end{vmatrix} = \frac{1}{F'_{Y|x}(y)}, \qquad (A7)$$

where $P_{Y|x}(y) = F'_{Y|x}(y)$.

Using equation (A2), we obtain that $N_Y$ is uniformly distributed on $[0, 1]$. If we assume that $N_Y$ is independent of $X$, then using the distribution transform theorem, we obtain

$$P_{Y,X} = \frac{P_X}{J} = P_X P_{Y|X}. \qquad (A8)$$

Figure 1

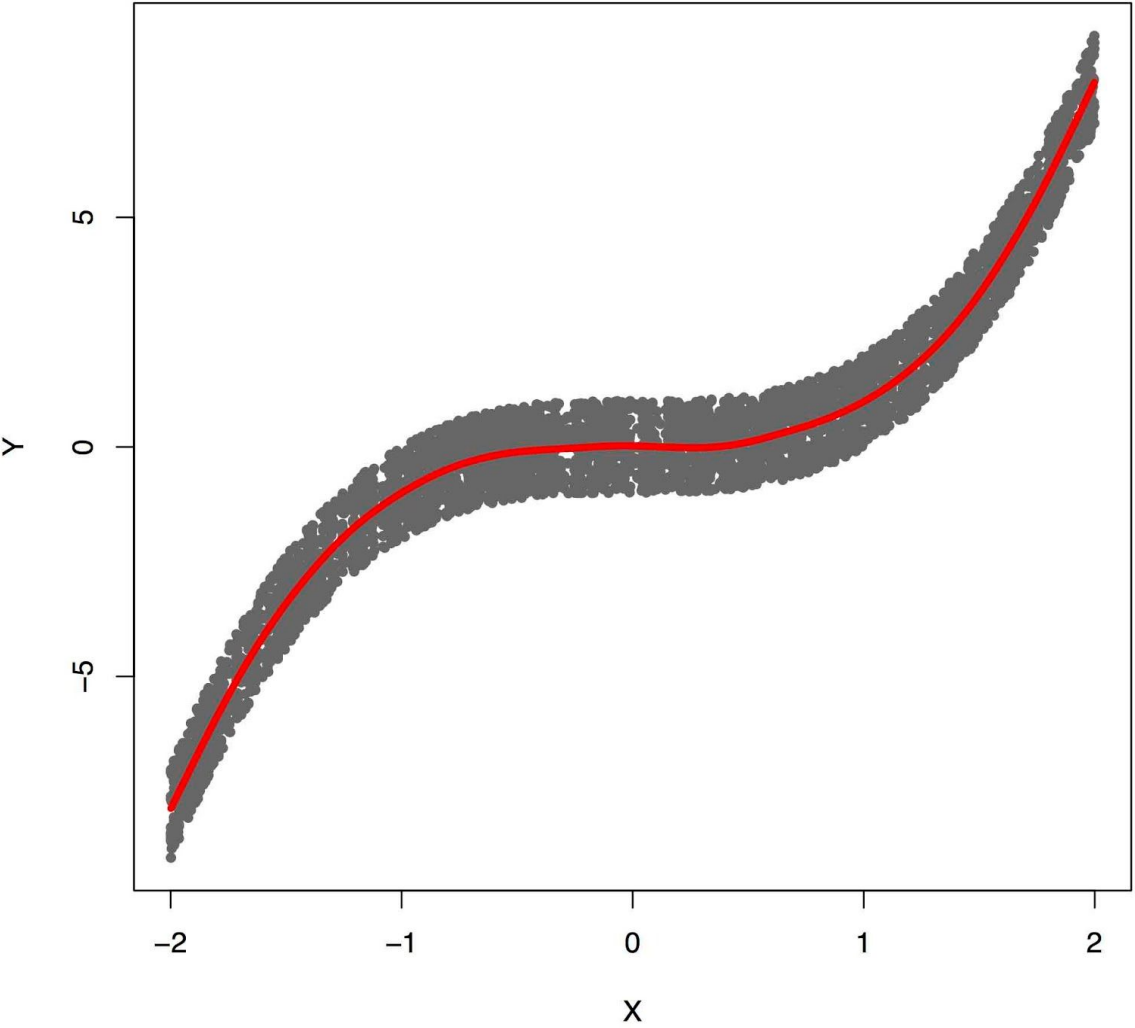

Figure 2

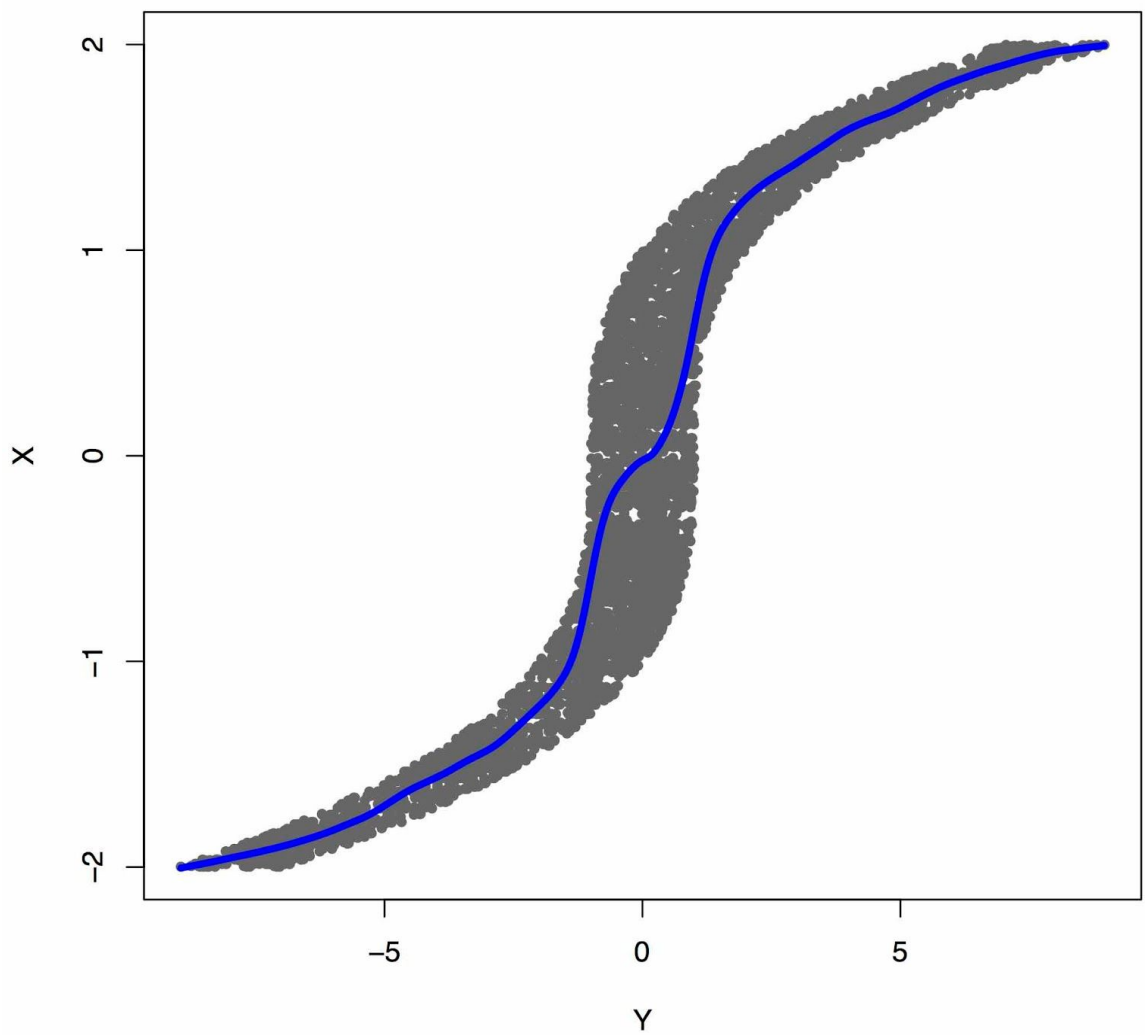

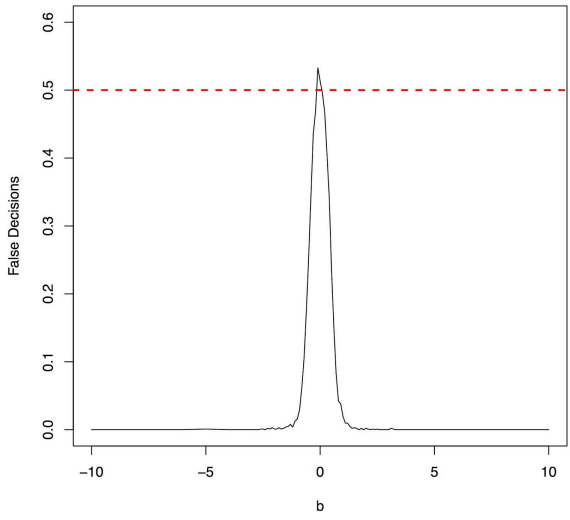

Figure 3

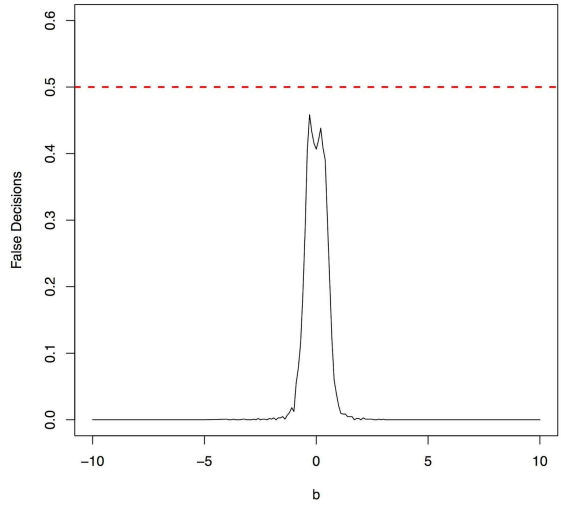

Figure 4

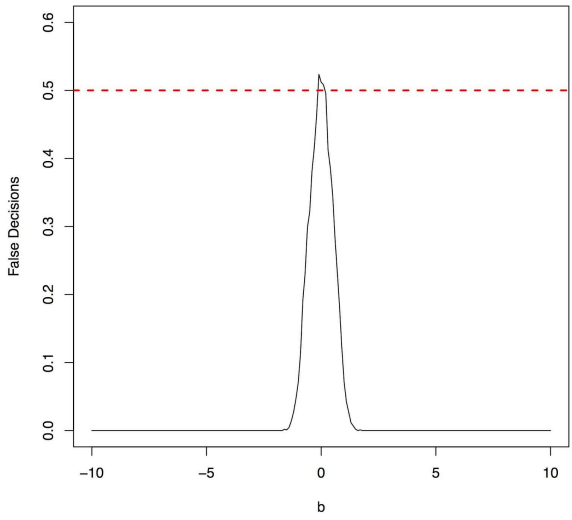

Figure 5

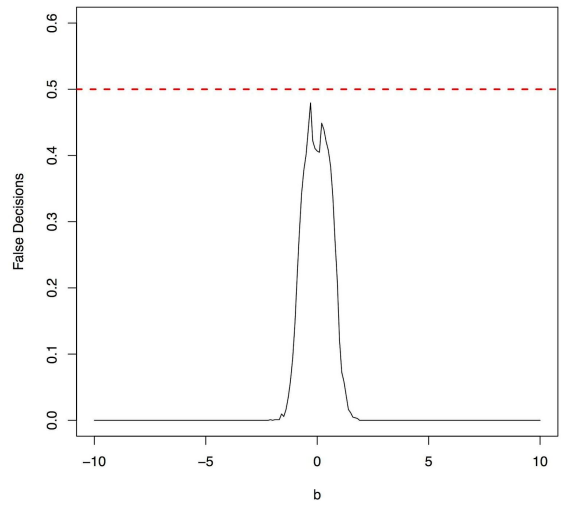

Figure 6

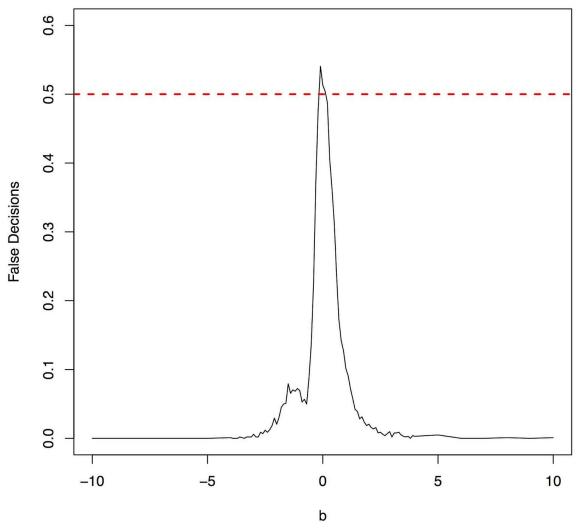

Figure 7

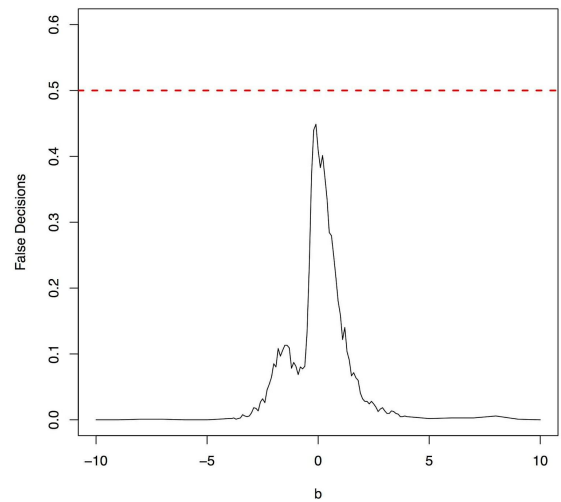

Figure 8

Figure 9

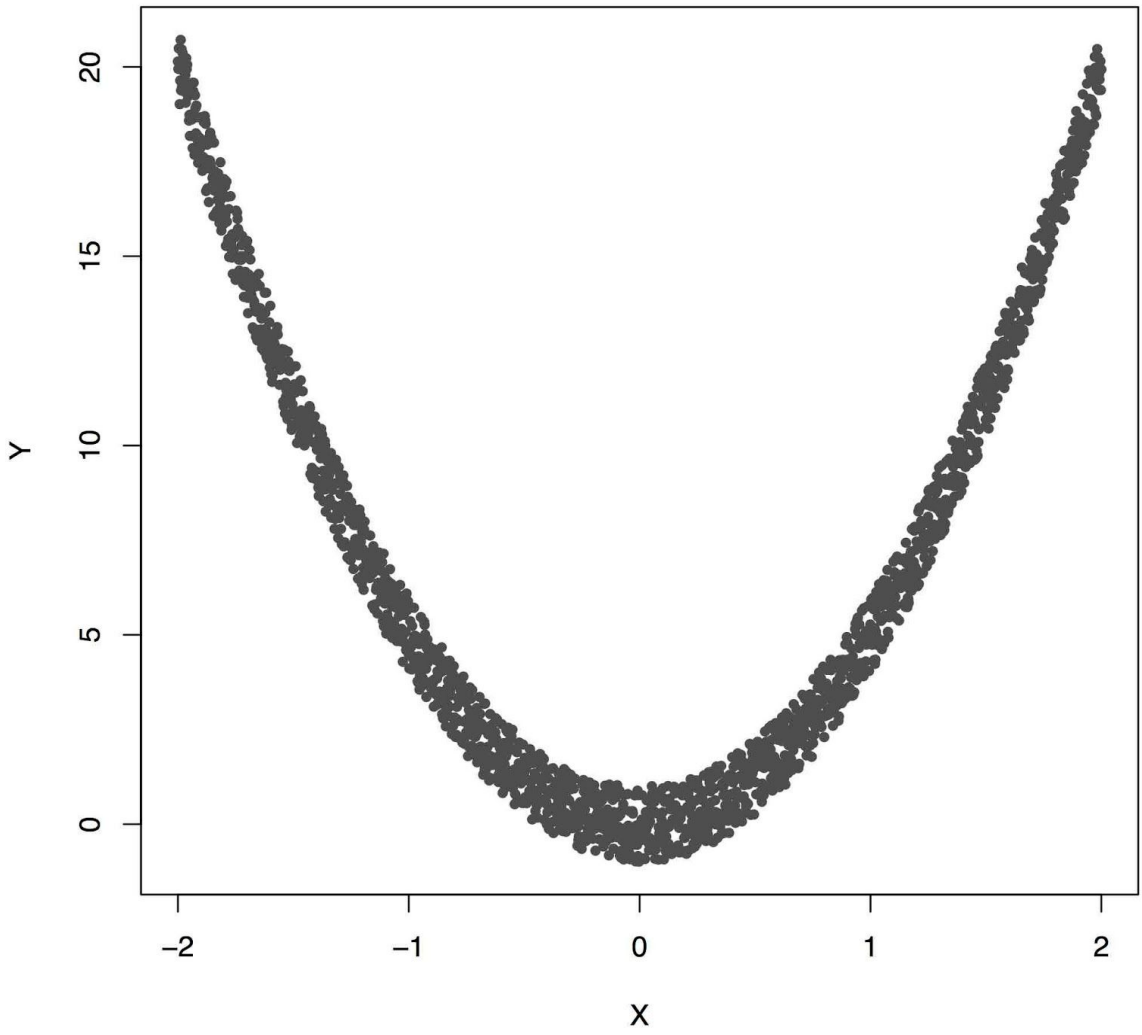

Figure 10

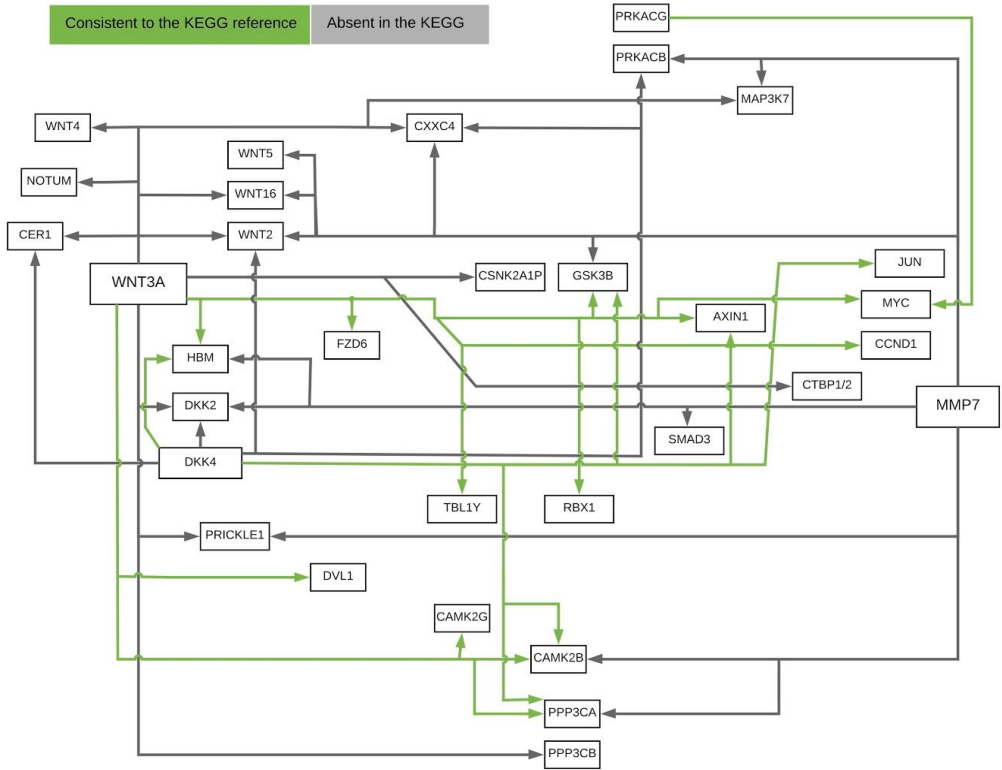

Figure 11

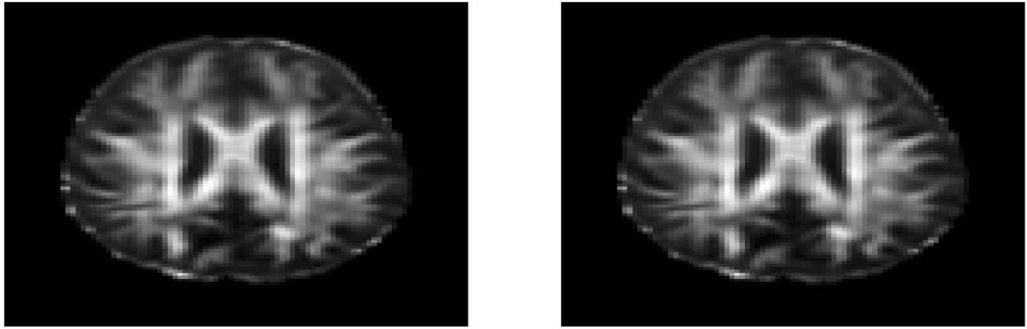

**(A)** A slice of the FA map from a single individual's DTI data.

**(B)** FA map reconstruction with the first two 3D-FPC scores.

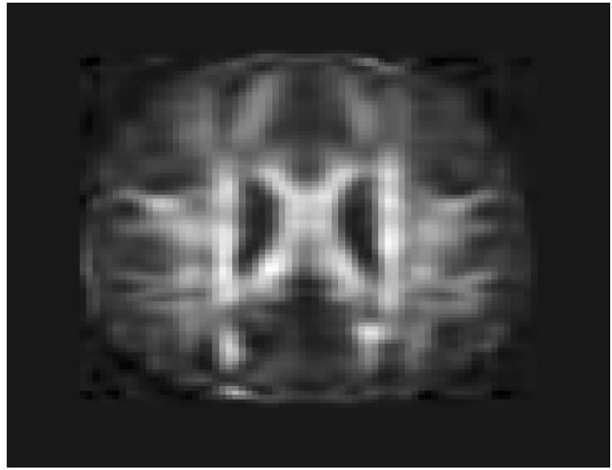

**Figure 12** Imputed FA map in Figure 11(A) using 3D-FPC scores and matrix completion.

Figure 13

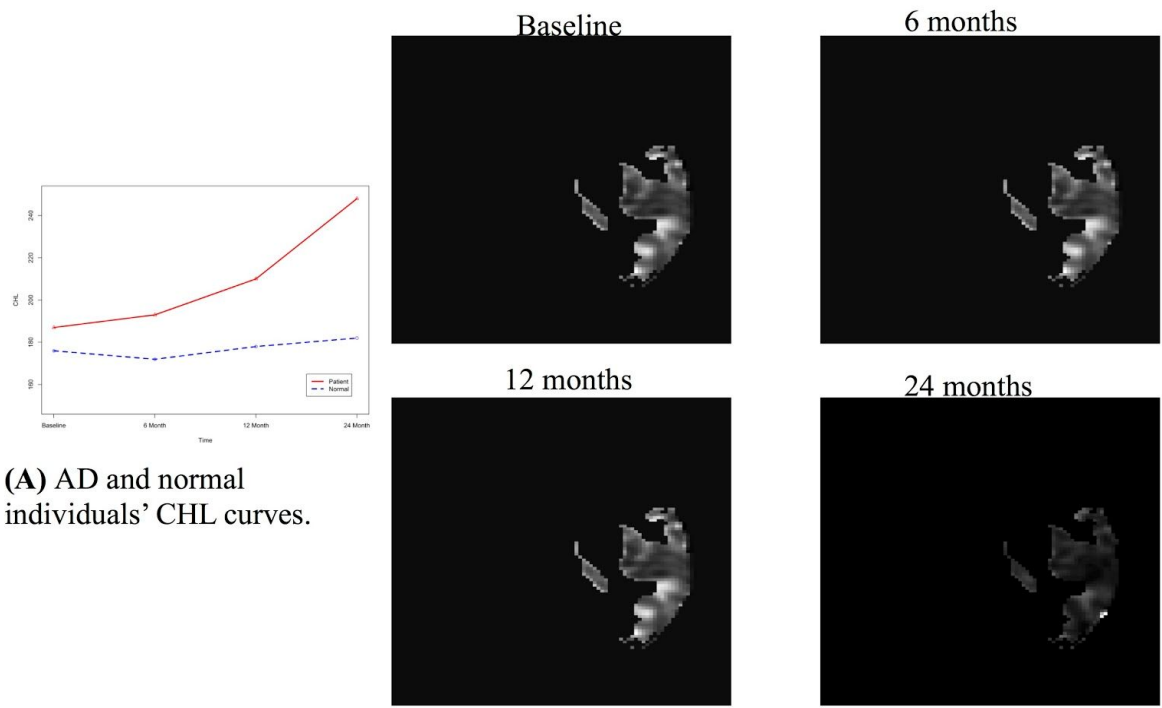

(A) AD and normal individuals' CHL curves.

(B) Images of temporal L hippocampus region.

Figure 14

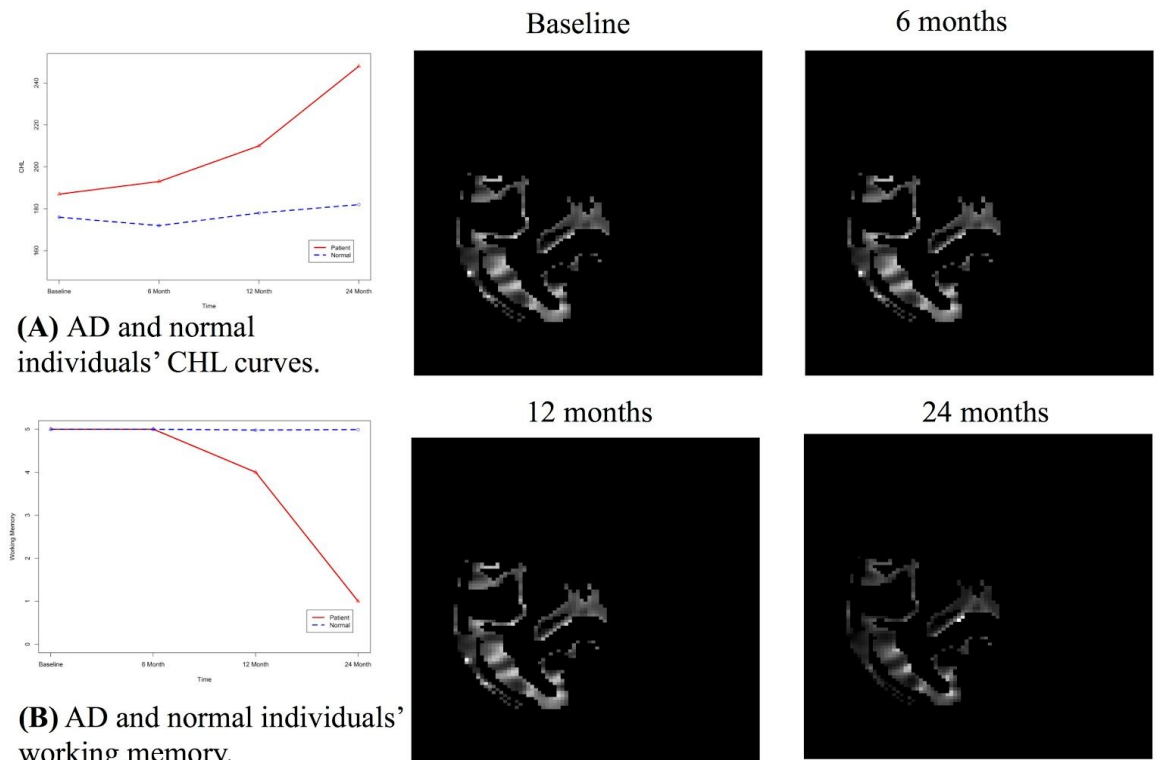

(A) AD and normal individuals' CHL curves.

(B) AD and normal individuals' working memory.

(C) Images of temporal R hippocampus region.

Figure 15

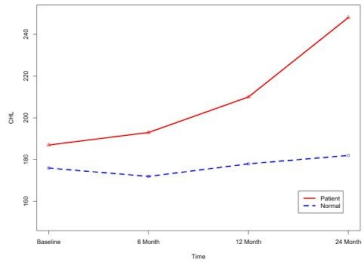

**(A)** AD and normal individuals' CHL curves.

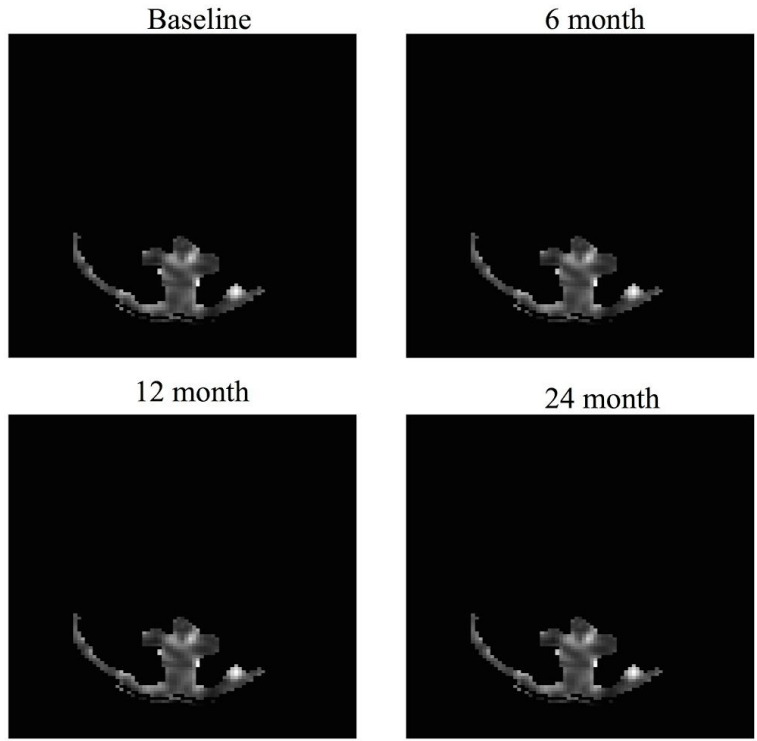

**(B)** Images of Occipital Lobe Region.